\documentclass[pdflatex,sn-nature]{sn-jnl}

\usepackage{graphicx}%
\usepackage{multirow}%
\usepackage{amsmath,amssymb,amsfonts}%
\usepackage{amsthm}%
\usepackage{mathrsfs}%
\usepackage[title]{appendix}%
\usepackage{xcolor}%
\usepackage{textcomp}%
\usepackage{manyfoot}%
\usepackage{booktabs}%
\usepackage{algorithm}%
\usepackage{algorithmicx}%
\usepackage{algpseudocode}%
\usepackage{listings}%
\usepackage{setspace}%
\usepackage[font={normalsize,doublespacing},labelfont=bf]{caption}
\usepackage{xr}
\begin{document}

\title[Article Title]{THz-induced phonon mode mixing and collective dynamics in a polar nanolattice}

\author*[1]{\fnm{Elizabeth} \sur{Skoropata}}\email{elizabeth.skoropata@psi.ch}
\author[2]{\fnm{Peter M.} \sur{Derlet}}\email{peter.derlet@psi.ch}
\author[1]{\fnm{Hiroki} \sur{Ueda}}
\author[1]{\fnm{Roman} \sur{Mankowsky}}
\author[1]{\fnm{Mathias} \sur{Sander}}
\author[1]{\fnm{Henrik T.} \sur{Lemke}}
\author[3]{\fnm{Rafael T.} \sur{Winkler}}
\author[1]{\fnm{Yunpei} \sur{Deng}}
\author[1]{\fnm{Milan} \sur{Radovic}}
\author[3]{\fnm{Matteo} \sur{Savoini}}
\author[1]{\fnm{Biaolong} \sur{Liu}}
\author[3]{\fnm{Martina} \sur{Basini}}
\author[3]{\fnm{Vladimir} \sur{Ovuka}}
\author[3]{\fnm{Steven L.} \sur{Johnson}}
\author[4]{\fnm{Marta D.} \sur{Rossell}}
\author[1]{\fnm{Urs} \sur{Staub}}\email{urs.staub@psi.ch}

\affil*[1]{\orgdiv{Center for Photon Science}, \orgname{Paul Scherrer Institute}, \orgaddress{\city{Villigen}, \postcode{5232}, \country{Switzerland}}}

\affil[2]{\orgdiv{Center for Scientific Computing, Theory and Data}, \orgname{Paul Scherrer Institute}, \orgaddress{\city{Villigen}, \postcode{5232}, \country{Switzerland}}}

\affil[3]{\orgdiv{Institute for Quantum Electronics, Physics Department}, \orgname{ETH Zurich}, \orgaddress{\city{Zurich}, \postcode{8092}, \country{Switzerland}}}

\affil[4]{\orgdiv{Electron Microscopy Center}, \orgname{Empa – Swiss Federal Laboratories for Materials Science and Technology}, \orgaddress{\city{D\"ubendorf}, \postcode{8600}, \country{Switzerland}}}

\maketitle

\textbf{Manipulating phonons through symmetry is a fundamental approach to alter the dynamic responses of materials. Most often new phases are sought through control of the unit-cell (e.g. via strain, doping, light, etc.). By comparison, there is vast potential to look beyond the unit-cell into higher-order architectures to control wave scattering and interference effects that remains less explored. We describe the THz-induced dynamics of an SrTiO$_3$ thin film with a nanoscale ordered interfacial dislocation network probed with 2- and 3-dimensional time-resolved x-ray diffraction and classical molecular dynamics simulations. We find that symmetry breaking at all scales is an effective approach to create a dynamical electric polarization and to control phonon mixing that generates previously unreported collective modes in the THz regime with circular vortex-like displacements. This work opens a new pathway to explore dynamical functional properties that can be extended to magnetic, electric, and ferroelectric systems by controlling the real-space topology via epitaxy. }

Symmetry is of fundamental importance to the properties of condensed matter systems, and its breaking or modification can lead to new phenomena and material functionalities. A well-known example is that of spontaneous symmetry breaking via an equilibrium phase transition, where variations in temperature or pressure give rise to new states of matter with distinctly different structural, electronic, and magnetic properties. Explicit symmetry breaking through micro-structural engineering or time-dependent perturbations that introduce new length- and time-scales can likewise drive material behavior. An exciting contemporary example is that of the two-dimensional Moir\'e lattice, where, through the control of translational symmetry via a twist angle, numerous exotic electronic, magnetic, photonic, and topological ground states emerge\cite{Wu_2017, Cao_2018, Wang_2019, Yankowitz_2019, Balents_2020, Shimazaki_2020, Song_2021, Xie_2023}. The same fundamental physics is also expressed in the dynamic regime, e.g., surface acoustic waves\cite{Yokouchi_2020, Nii_2025}, phononic crystals\cite{Graczykowski_2015, Ma_2019, Tanaka_2000, Florez_2022}, and transient gratings\cite{Bencivenga_2015, Rouxel_2021} that combine spatially varying local order parameters and long-range periodicity to control the nonequilibrium lattice response and wave propagation. 

In strongly correlated electron systems, an exemplar periodic nanostructure is the ferroelectric/dielectric superlattice PbTiO$_3$/SrTiO$_3$ (PTO/STO) with 3-dimensional polar textures\cite{Yadav_2016}. These systems host intriguing collective modes and demonstrate the ability to manipulate dielectric responses on ultra-fast timescales using light or electric fields\cite{Stoica_2019}. These systems were also found to stabilize chiral topological structures\cite{Das_2019} with a potential to surpass the performance of analogous magnetic solitons in information applications by exceeding the speed of magnetic bits\cite{Li_2021, Wang_2025, Li_2025}. Understanding the coupling mechanisms between external light and functional degrees of freedom, such as ferroelectricity and magnetism, when modulated by a nanostructure will potentially enable the use of collective modes for future ultra-fast electronics. The discovery of solitonic states in polar vortex superlattices and predicted in other systems (e.g. twisted/rumpled/surface states of STO\cite{Xu_2023, Xu_2024, Sha_2024, Tenreiro_2026} and twisted BaTiO$_3$\cite{S_nchez_Santolino_2024}) highlights the close connection between local polarization and atomic displacement, and suggests the potential for a rich landscape of dynamical phenomena. In STO in particular, local inhomogeneities can hybridize phonon modes to couple polar optical and acoustic responses. Despite this, a direct atomic-scale link to the underlying symmetry breaking responsible remains unclear\cite{Tenne_2007, Evarestov_2012, Hameed_2021, Fauqu__2022, Orenstein_2025}. Here, we address how THz-induced phonons scatter from an ordered nanolattice in STO leading to phonon mixing and temporally evolving strain fields to create novel collective modes and a dynamical electric polarization. 

Recently, we have created and investigated a new nanostructure topology in epitaxially grown SrTiO$_3$ thin films on a (LaAlO$_3$)$_{0.3}$(Sr$_2$TaAlO$_6$)$_{0.7}$ (LSAT) substrate. After a high temperature annealing process, the lattice mismatch of the STO and the LSAT is relaxed through the formation of a two dimensional misfit dislocation network\cite{Burian_2021} with similar periodicity as the observed polar vortex structure in the PbTiO$_3$/SrTiO$_3$ (PTO/STO) multilayers\cite{Yadav_2016} and photoinduced supercrystals\cite{Stoica_2019}. Here, we report on ultrafast pump-probe experiments in which THz radiation interacts with the charged atoms in the STO thin film and where symmetry breaking creates a mixture of sub-THz longitudinal and transverse acoustic modes. Our unique approach to generate mixed phonon modes through controlled local and nanoscale symmetry breaking drives a new dynamic phase wherein linear (in-plane and out-of-plane) motions mix to create circular vortex-like atomic collective modes and a dynamic electric polarization on timescales comparable to those recently observed in ferroelectric superlattices\cite{Li_2021}. 

\subsection*{Sub-THz dynamics of a periodic nanostructure from 2- and 3-dimensional time-resolved x-ray diffraction}
We used a single-cycle THz excitation and time-resolved hard x-ray diffraction (tr-XRD) probe at the Bernina endstation of the Swiss Free Electron Laser (SwissFEL)\cite{Ingold_2019, Mankowsky_2021} on a 34 nm-thick film of STO grown on a (001)-oriented LSAT substrate\cite{methods} cooled to 5~K. An interfacial network of misfit-dislocation lines with a 40~nm spacing forms a square grid at the film-substrate interface along the $a$ and $b$ cubic crystal axes as shown in (Fig.~\ref{fig:Fig1}a-b) and Ref.\cite{Burian_2021}. The atomic displacements around the dislocations result in a local inhomogeneous strain (e.g. Supplementary Fig.~1). The high degree of periodicity of the dislocation network is visible in XRD as superstructure satellites that are well resolved in momentum space. A strong single-cycle (600 kV/cm) THz pulse (Supplementary Fig.~2) with vertical polarization was applied parallel to the [1-10] direction of the STO/LSAT. The THz electric field interacts with the charged atoms through the soft mode resulting in a polar displacement within the STO film\cite{Li_2019, Kozina_2019}.  

To probe the structural response of the system we monitored the intensity of the (1-13) reflection on an area detector and integrated regions corresponding to the maximum intensity of the STO crystal structure Bragg reflection (Fig.~\ref{fig:Fig1}c) and representative dislocation network satellite reflections with ($H_{\rm{s}}$ $K_{\rm{s}}$) = (01), (11), and (10) (S1, S2, S3) indexed by the 40~nm supercell. The transient response of the diffraction intensity $\Delta I/I = (I_{\rm{on}} - I_{\rm{off}})/I_{\rm{off}}$ occurs over picosecond time scales. The fast Fourier transform (FFT) of the STO reflection time trace consists of two main modes at $\nu = $~0.060 and 0.124~THz. The superstructure reflections show the same modes and include additional frequency components above $0.15$~THz. The timescale of the observed dynamics are slower than the polar 1.2~THz soft-mode of STO\cite{Marsik_2016, Li_2019, Kozina_2019}. We first discuss the response of the STO Bragg reflection where the modes are within the frequency range of coherent acoustic phonons (strain waves)\cite{Mattern_2023}. Along the cubic axes of STO the longitudinal and transverse speeds of sound are $v_L$ = 7.9 and $v_T$ = 4.9 nm/ps, respectively\cite{Bell_1963}. The 0.124 THz frequency matches a longitudinal strain wave completing a full cycle through the film thickness $\nu = v_L \times 2t_{\rm{film}}$. The additional lower frequency mode that we observe suggests a more complex dynamics that may be considered a mixture of longitudinal and transverse acoustic waves or other collective modes as observed in ferroelectric superlattices\cite{Li_2021, Wang_2025}. We also note that while the THz pulse (Supplementary Fig.~2) has a duration of 2~ps there is a delay in the largest pump effect occurring at 20~ps suggesting that wave scattering and interference effects contribute to the maximal lattice response. The additional higher frequencies observed at the dislocation superstructure reflections are too fast to be described by simple coherent acoustic phonons/strain waves considering the speed of sound and the length scales of the film or dislocation network periodicity. However, our system of misfit dislocations, characterized by a periodic inhomogeneous strain field including transverse, shear, and longitudinal components, provides a pathway for longitudinal and transverse acoustic modes to mix via scattering to create new collective modes. The local inversion symmetry breaking of the dislocation may also create polar fluctuations and hybrid optical-acoustic phonons, as observed in inelastic neutron scattering\cite{Fauqu__2022} and diffuse x-ray scattering with THz excitation\cite{Orenstein_2025} (discussed further below).

Examining the time dependence of pumped--unpumped differences of the 2-dimensional detector images centered on the (1-13) STO Bragg reflection (Fig.~\ref{fig:Fig2}a-b) give more insight to the type of atomic motions generated by the THz excitation. A clear contrast ($I_{\rm{on}} - I_{\rm{off}}$) in the STO reflection in the out-of-plane $L$ direction is observed when the transient response is largest (e.g. the maxima at $t = 3.0$~ps and $t = 19.2$~ps in the time traces of Fig.~\ref{fig:Fig1}). When the STO reflection response is small, an in-plane ($H,K$) contrast in the satellite reflections is observed. This indicates an exchange of out-of-plane motion and in-plane modulation at the dislocations similar to domain walls\cite{Locherer_1998}, in-plane gradients switching polar nanoregions of relaxor ferroelectrics\cite{Xu_2006}, or as in polar textures of ferroelectric superlattices\cite{Zatterin_2024}.

To better understand the dynamics indicated by the tr-XRD, we measured the 3-dimensional reciprocal space volume (RSV) surrounding the (1-13) STO reflection at selected pump-probe delay times as shown in Fig.~\ref{fig:Fig2}c-d. At $t = 3.0$~ps and 19.2~ps the signal contrast has a strong component along the $L$ direction for all parts of the RSV indicating that an average out-of-plane lattice expansion accompanies the transient state at those time points. By comparison, at 8.6 and 13.0~ps, where the change in total diffraction intensity is weaker, there is an exchange of intensity between the $\pm$(10) and $\pm$(01) satellite reflection pairs. At the x-ray photon energy used, the imaginary component of the x-ray scattering factors is large enough to make the intensity among the superstructure pairs unequal under a polar displacement. Therefore, the observed exchange of intensity in superstructure reflection pairs (distinct from a simple shift in reflection position in reciprocal space) indicates a breaking of the local symmetry of the dislocation lattice due to a dynamic modulation of the electric polarization\cite{Fujimoto_1978, Gorfman_2016}. We use a line cut along $L$ at the STO Bragg reflection to estimate the transient change in the average out-of-plane ($c$-axis) lattice parameter vs. time for the first 25~ps (Supplementary Fig.~3) and find that the average lattice expansion has a period that matches that of the lowest frequency mode of the STO reflection from the time trace. The limited time points of collected RSVs hinders an estimation of the time scale of the superimposed in-plane motions and polar dynamics directly from 3-dimensional reciprocal space; however, a careful examination of the 2-dimensional images shows clearly that there is a persistent mixture of in-plane and out-of-plane motions, even when the out-of-plane lattice motion appears dominant. 

\subsection*{Model simulations of THz-driven dynamics of SrTiO$_3$ thin films}
From the tr-XRD, we observe a mixture of sound waves and a local polarization emerging from the ordered dislocation network with sensitivity to the macroscopic behavior of the system, but the scattering signals are insufficient to directly determine the internal atomic-scale motions of the observed modes. To explore this more deeply we used classical molecular dynamics simulations employing an empirical core-shell model analagous to the model used to study polar vortex structures in STO\cite{Li_2021}. This bottom-up approach considers the force constants and electric charges of each ion, and allows a study of the response of a polarizable system to a perturbing electric field, from which the time-dependent atomic motions and phonons emerge as a result\cite{Sepliarsky_2005}. Two model STO geometries are considered, one containing an interfacial misfit [100]-[010] dislocation network and one with a pristine thin film free of dislocations\cite{methods}. To simulate the entire pump-probe experiment, we first equilibrate our systems at room temperature and then apply a time dependent in-plane electric field whose waveform is identical to that of the experiment. To benchmark the calculations, we first make a direct comparison to the experimental tr-XRD data by calculating the time dependent scattering intensities in the vicinity of the (1-13) reflection from the simulated atomic positions. Figure~\ref{fig:Fig1}e shows the Fourier spectrum of the changes in the calculated STO and superstructure satellite reflection intensities revealing qualitatively similar mode frequencies as obtained from the tr-XRD data after scaling the simulated cell and film thickness to the actual dislocation network and film thickness spacing\cite{methods}. The RSVs from the molecular dynamics simulations shown in Supplementary Fig.~7 also reveal the same time-dependent contrast in the superstructure satellite intensity further supporting the existence of dynamically modulated electric polarization of the dislocations.

A comparison of the atomic motions for the dislocation model system with a pristine geometry with dislocations further reveals the nature of the observed modes. During the THz pulse both systems show in-plane atomic displacements with a transient electric polarization. Interestingly, we find that the in-plane motion directly driven by the THz couples strongly to out-of-plane motion at the surface of the film during the THz pulse (Supplementary Fig.~9). This is likely connected to the simultaneous observation of the acoustic and ferroelectric soft modes in THz pump diffuse x-ray scattering experiments probing at nonzero momentum transfer\cite{Orenstein_2025} which indicates a strong coupling of optical and acoustic modes in symmetry-broken points within STO. Also, while recent theory work indicated a THz excitation can produce a bulk out-of-plane expansion\cite{Libbi_2025}, our findings show that, in a thin film, the surface is a key source of inversion symmetry breaking that connects the optical and acoustic modes.

Following the THz pulse, in-plane and out-of-plane motions are both present, generating different types of dynamical atomic displacements and strain. As shown in Fig.~\ref{fig:Fig3}a-b, immediately following the THz pulse, the pristine STO thin film displays a homogeneous lateral and inhomogeneous depth-dependent response with a 2-dimensional character resembling a superposition of in-plane motion initiated by the THz field with an out-of-plane strain wave. For STO with dislocations, a further in-plane symmetry breaking transforms the response into 3-dimensions; a mixture of in-plane and out-of-plane displacements in the [111] and [-1-1-1] directions in a checkerboard pattern in the deepest part of the film is imposed by the periodic nanostructure. A visual inspection of the atomic displacements in 3-dimensions (Fig.~\ref{fig:Fig3}c) shows that the presence of dislocations generates a circular motion not observed in the pristine system.

\subsection*{Polarization and vorticity of the sub-THz modes in STO nanostructures}
\begin{figure} 
	\centering
	\includegraphics[width=0.7\textwidth]{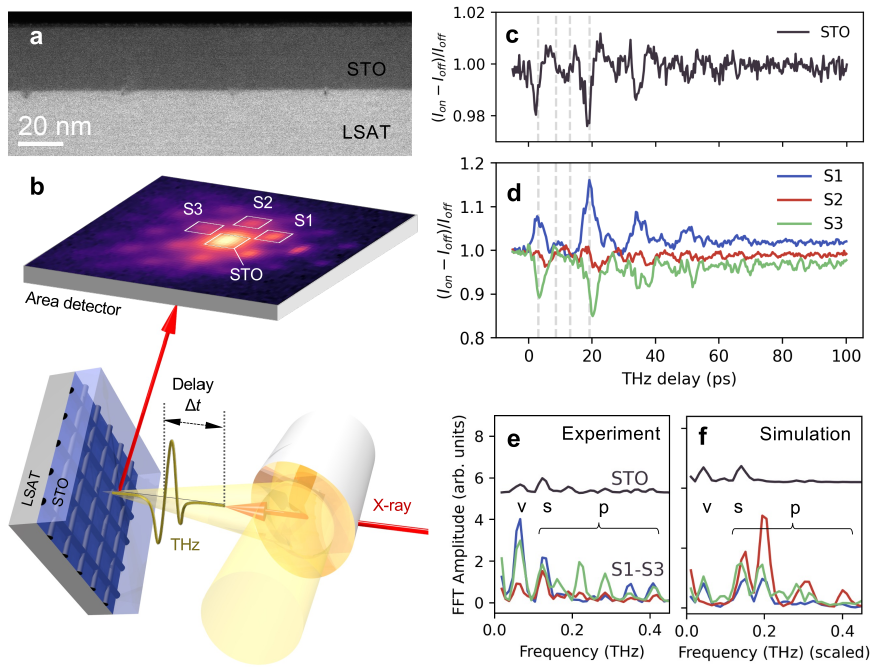} 
	\caption{\textbf{Collective modes created in a polar nanostructure}: \textbf{a} Cross-sectional scanning transmission electron microscopy (STEM) of the STO/LSAT revealing a periodically ordered array of misfit dislocations at the interface. \textbf{b} Schematic of the THz-pump x-ray diffraction probe experiment using an x-ray free electron laser with a typical area detector image identifying the STO (1-13) reflection and its satellites due to nanostructure order. \textbf{c} Time-dependent diffraction intensity of the (1-13) reflection \textbf{d} and its nanostructure satellites, and \textbf{e} Fourier spectra of of the time-dependent response of the bulk STO and nanostructure from transient x-ray diffraction compared with \textbf{f} the simulated response from classical molecular dynamics simulations. The labels $v$, $s$, $p$ indicate the characteristics of the modes (vorticity, strain wave, and polar, respectively) as described in the Main Text.}
	\label{fig:Fig1} 
\end{figure}

\begin{figure}[t!]
	\centering
	\includegraphics[scale = 0.65]{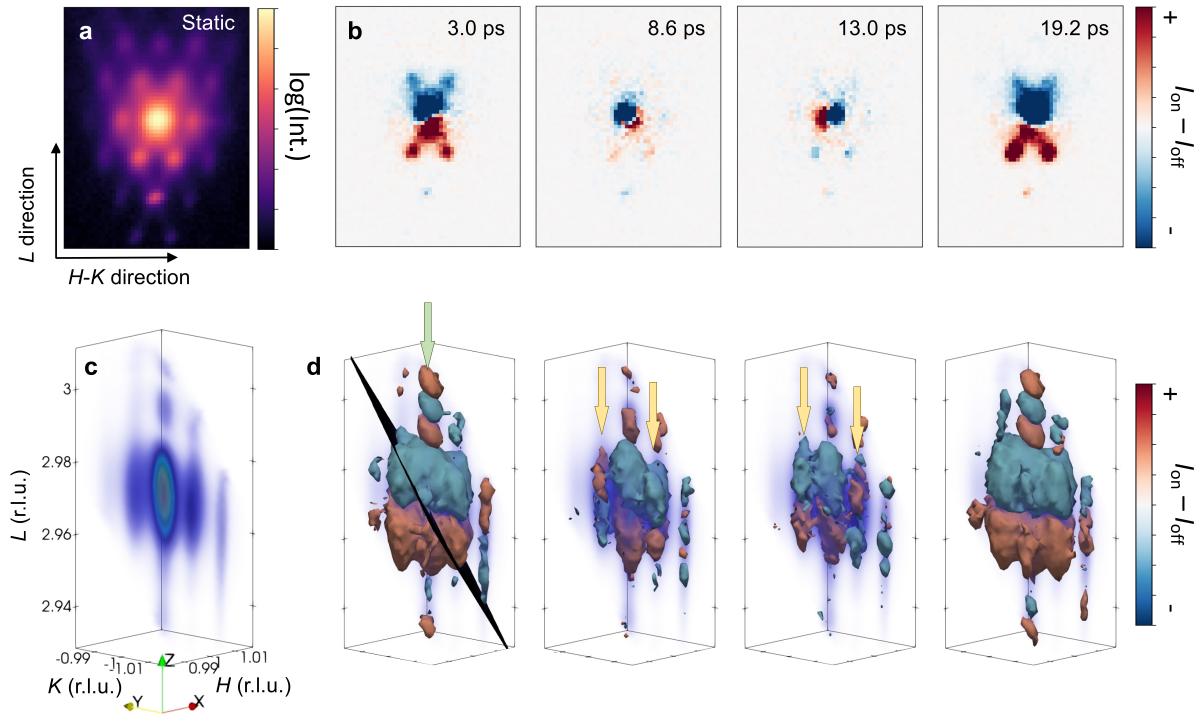}
	\caption{\textbf{Time-dependent changes of structure and polarization in 2- and 3-dimensional reciprocal space}:  \textbf{a} Static and \textbf{b} pump-probe 2-dimensional reciprocal-space planes (black plane in d) centered on the film STO reflection. \textbf{c} Static reciprocal-space volume and \textbf{d} pump-probe 3-dimensional reciprocal-space volume isosurface surrounding the (1-13) reflection of SrTiO$_3$ with an interfacial dislocation network. The time points are indicated with the dashed lines in Fig.~\ref{fig:Fig1}c-d}
	\label{fig:Fig2}
\end{figure}

\begin{figure}[h!]
	\centering
	\includegraphics[scale = 0.7]{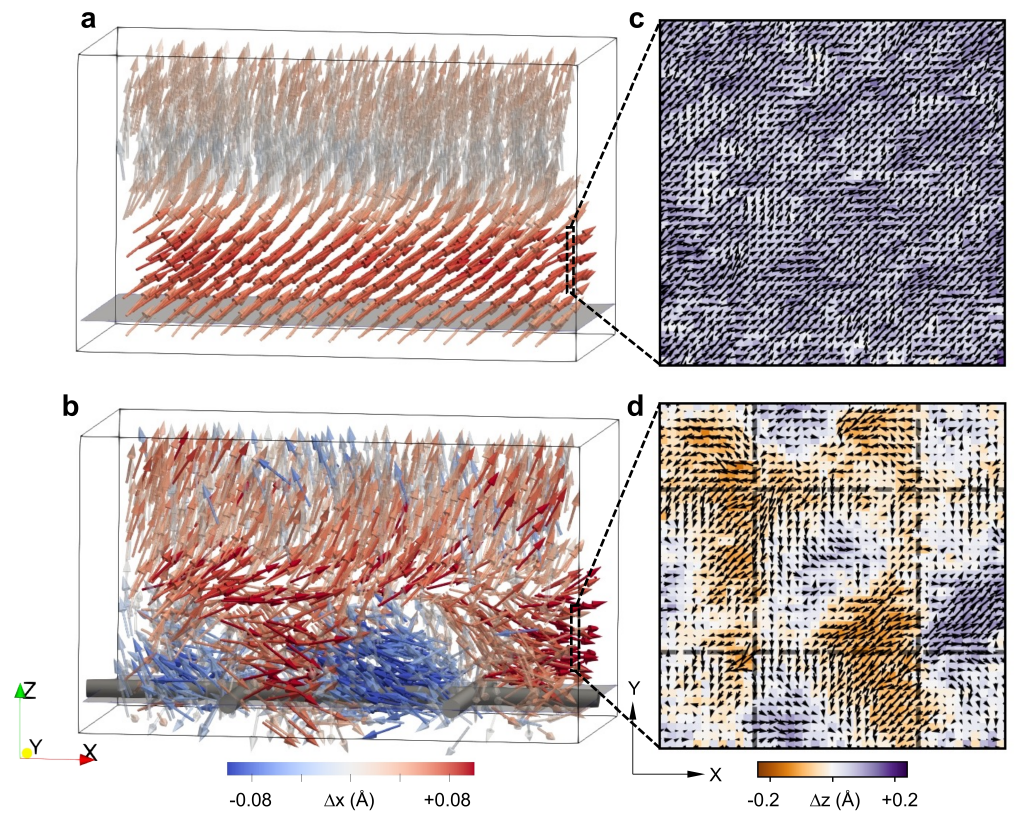}
	\caption{\textbf{Transformation from 2-dimensional waves to 3-dimensional vortex-like displacements in SrTiO$_3$ thin films}: \textbf{a}, \textbf{b} 3-dimensional and \textbf{c}, \textbf{d} 2-dimensional in-plane illustration of the O$^{2-}$ displacement (averaged over 3 layers in the depth, $z$, indicated by the dashed box) immediately following the THz pulse ($t_{\rm{off}}$ in Supplementary Fig.~2) for pristine (upper panels) and dislocation network STO films (lower panels). The colour in (a, b) indicates the in-plane displacement magnitude and the vector indicates the displacement direction (same magnitude used for simplicity of visualization). The colour in (c, d) indicates the out-of-plane displacement magnitude and the vectors are proportional to the x-y displacement direction and magnitude. Animations of the O$^{2-}$ displacements vs. time are provided in the Supplementary Materials.}
	\label{fig:Fig3}
\end{figure}

\begin{figure}[h!]
	\centering
	\includegraphics[scale = 0.6]{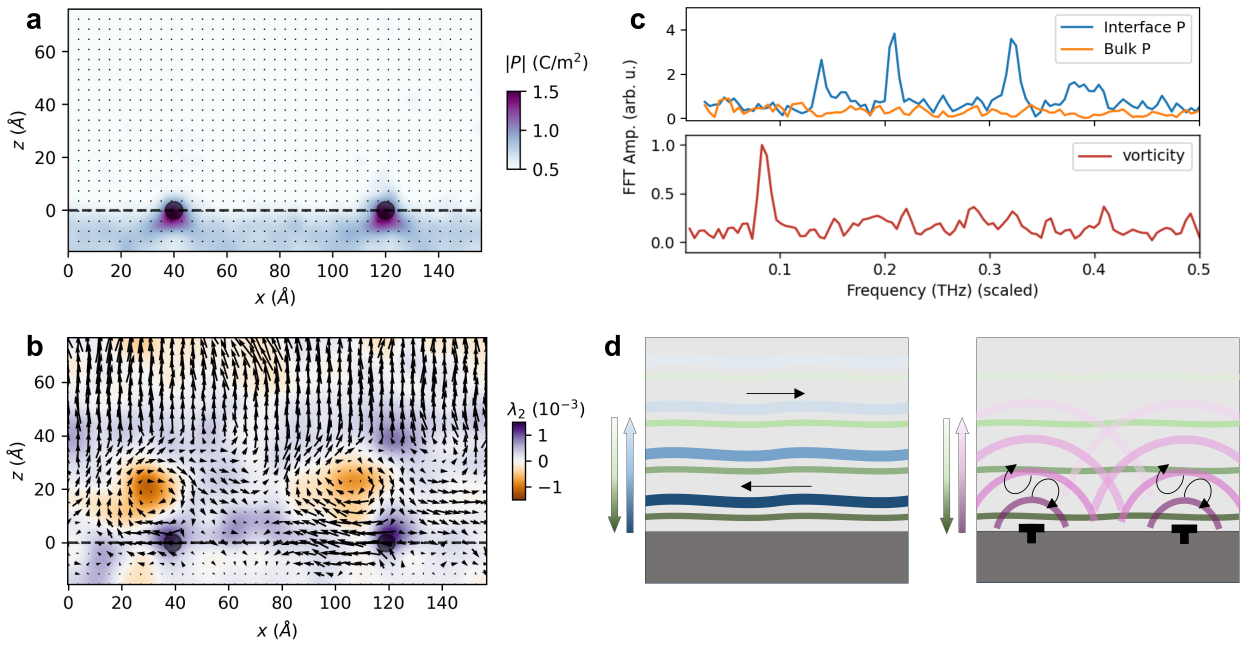}
	\caption{\textbf{Polarization and vorticity of phonon modes in nanolattice SrTiO$_3$}: 2-dimensional projections of \textbf{a} the magnitude of the static electric polarization and \textbf{b} the $\lambda_2$ vorticity criterion in STO with an interfacial dislocation network immediately after a THz pulse (time point $\#$3 in Supplementary Fig.~2). The black arrows in (A, B) indicate the O$^{2-}$ displacement. \textbf{c} Fourier transform of the average polarization (details in Supplementary Fig.~11), and the time-dependent film vortcity from classical molecular dynamic simulations. \textbf{d} Illustration of the transformation of the dynamics of STO thin films from pure wave-like to turbulent vortex-like arising from a periodic interfacial dislocation network.}
	\label{fig:Fig4}
\end{figure}

We now address the mode dependent atomic-level transient polarization $|P|$ as a function of time. The atomic polarization vector, derived from the simulated atomic motion, was calculated by summing up the charge-weighted radial distance vectors of each nearest neighbor environment --- nonzero values will originate from thermal vibrations and from the static broken symmetries due to the dislocations and surface, and the dynamical disturbances due to the THz pump. The small static polarization due to the dislocation network can be seen in Fig.~\ref{fig:Fig4}a and Supplementary Fig.~10 both before and after the THz excitation. The mode selective changes in $|P|$ are obtained from the FFT spectrum of the time-dependence of $|P|$ averaged among the interfacial layers along the dislocation lines and layers closer to the bulk-like region as well as for the pristine film. No polar dynamics are observed after the THz pulse in the pristine film or in the bulk-like region of the film with dislocations (Fig.~\ref{fig:Fig4}c). Polar dynamics are, however, seen in the interfacial region containing the dislocations (Supplementary Fig.~11). While the underlying displacements are fundamentally elastic (described by constitutive equations not involving non-linear advection) the spatially heterogeneous polarization suggests turbulent-like fluctuations. Furthermore, regions away from the dislocation structure and the pristine structure in its entirety, exhibit laminal-like displacements, where the displacement fields are predominantly homogeneous. We quantify the laminar-like vs. turbulent-like behavior of the displacement patterns using methods developed by the fluid dynamics community to identify their level of vorticity. We consider our displacement field analogously to a velocity field ($\mathbf{u}$) of a fluid and using: 1) the local helicity parameter $H$\cite{Moffatt_1969} which measures the local helicity via $H_{i}=\mathbf{u}\cdot(\nabla \times \mathbf{u}$)) and 2) the $\lambda_{2}$ parameter\cite{Jeong_1995,Otto_2012}, which identifies regions where rotation dominates over strain (when $\lambda_{2}<0$). Specifically $\lambda_{2}$ is the second largest eigenvalue of the matrix $\mathbf{S}^2 + \mathbf{\Omega}^2$ with the local rate of strain $\mathbf{S} = \frac{1}{2} \left( \nabla \mathbf{u} + (\nabla \mathbf{u})^\mathrm{T} \right)$ and the local rotation $\mathbf{\Omega} = \frac{1}{2} \left( \nabla \mathbf{u} - (\nabla \mathbf{u})^\mathrm{T} \right)$. Both measures are used as a combined ``vorticity" criterium since regions with minimal collective displacements are strongly affected by the small displacements arising from thermal noise. The FFT spectra of the time dependence of the vorticity simulation data ($\lambda_{2}$) is shown in the lower panel of Fig.~\ref{fig:Fig4}c. These findings allow us to classify the modes based on the details of the 2- and 3-dimensional experimental tr-XRD and atomic displacements. They are associated with a (predominant) longitudinal strain wave (s), a vortex-like atomic displacements with transverse character (v) or mixed longitudinal and transverse acoustic and modes with a 3-dimensional character that modulate the polarization of the nanostructure (p), as labeled in (Fig.~\ref{fig:Fig1}e-f).

\subsection*{Creation of mixed phonon modes in SrTiO$_3$ via symmetry breaking}
A key finding of ours is that the THz excitation of an in-plane optical mode generates in-plane and out-of-plane motion with longitudinal and shear strain in a pristine STO thin film (Supplementary Fig.~12). With a dislocation network, the shear modes generated by the THz become strongly scattered and drive new transverse modes (Supplementary Fig.~13). This results not only in a local symmetry breaking at the dislocation core, but a mesoscopic transient symmetry breaking of the dislocation network lattice that fundamentally transforms the lattice response vs. the pristine film. Indeed, as dislocations themselves are robust topological defects characterized by anisotropic strain, they hold promise to transform polar textures, e.g., as explored in a recent theoretical effort to manipulate polar vortices through engineering model screw dislocation structures\cite{Chen_2025}. The analogy with fluid turbulence, and more specifically, between displacement field and fluid flow\cite{Chen_2022} is beneficial to provide a framework to describe the phonon modes resulting from our periodic nanolattice of dislocations. For our model dislocation-free pristine system laterally homogeneous strain without evidence of a vortex structure is observed. This contrasts with the interfacial dislocation geometry, where phonon scattering leads to heterogeneous strain at the scale of the dislocation structures and ultimately to the emergence of turbulent-like displacement structures with quantifiable vorticity properties. In bulk STO, recent first-principles quantum nuclear dynamics calculations indicated that in-plane THz pulses can induce out-of-plane stresses\cite{Libbi_2025}. Our result is consistent with this finding but further demonstrates that the symmetry breaking of a thin film surface and dislocation network are key to mixing the acoustic/dynamic strain modes of the system and produce collective dynamics similar to the THz-driven collective modes in polar vortex nanostructures\cite{Li_2021, Wang_2025, Li_2025}. 

We also observe clear polar dynamics occurring on timescales faster than acoustic modes within the STO dislocation nanostructure both in our experimental tr-XRD and in simulation. Microscopy studies of dislocation cores and point defects in STO have measured sizeable local strain fields resulting in anisotropic ferroelectric order and a local polarization\cite{Gao_2018, Li_2021b, Sha_2023}. In the dynamic regime, polar nanodomains, defects, and local inhomogeneities in STO can lead to new phonon modes and modify or hybridize optical and acoustic modes \cite{Tenne_2007, Evarestov_2012, Hameed_2021, Fauqu__2022, Orenstein_2025}, but a direct simultaneous measurement of local polarization and macroscopic phonon responses has remained a challenge. Our molecular dynamics simulations support the existence of a static polarization in an STO dislocation network and our tr-XRD experiment probes the dynamically varying polarization to give a first clear macroscopically sensitive confirmation of inhomogeneity-induced local polarization and its dynamic modulation with strain. 

Our approach to symmetry breaking in STO that couples polar and the acoustic lattice responses sheds new light to the phonon mode hybridization that has been a key open question regarding the nature of the quantum polar instability.  Moreover, phonon mixing from mesoscopic length scales using our nanostrcucture design principle opens a new pathway to explore symmetry-driven states in both space and time. This general approach extended to other systems may greatly expand the potential to control transient collective states or real-space charge, spin, lattice, or orbital topologies such as chiral objects, spin or charge skyrmions, polar vortices, or recently discussed ferrons, and paves the way for reconfigurable quantum, electronic, thermal, or spintronic functionalties across diverse materials classes.  

\clearpage

\clearpage
\singlespacing
\bibliography{STO_MD_dynamics}

\doublespacing
\section*{Methods}

\subsubsection*{Sample growth}
A SrTiO$_3$ film was grown on LSAT using pulsed laser deposition, and is the same as used in the experiments\cite{Kozina_2019, Burian_2021}. To create the interfacial dislocation network, the sample was annealed in air for 12~h at 1200$^{\circ}$C. The thickness was measured using x-ray reflectometry using a Bruker D8 Discovery diffractometer with Cu K$_{\alpha}$ x-rays and the thickness and roughness modeled using GenX\cite{Glavic_2022}. The reciprocal space volume (RSV) surrounding the (1-13) reflection was measured at the Surface Diffraction endstation of the Material Science beamline at the Swiss Light Source\cite{Willmott_2013} where the UB matrix was defined relative to the LSAT substrate. The RSV was reconstructed using xrayutilitles\cite{Kriegner_2013} and indicted a 2-dimensional ordered grid in the in-plane $a$- $b$-directions with a spacing of 40 nm between the dislocation lines. 

\subsubsection*{Transmission electron microscopy}
Electron-transparent samples for STEM investigations were produced in a cross-section geometry using an FEI Helios 660 G3 UC dual-beam focused (Ga) ion beam instrument operated at 30 and 5~kV, after the deposition of C and Pt protective layers. Scanning transmission electron microscopy (STEM) was performed using a probe aberration–corrected FEI Titan Themis microscope operated at 300~kV. Data were acquired with a probe convergence semiangle of 26~mrad  in combination with a collection angular range for the high-angle annular dark field (HAADF) detector set to 90–170 mrad. Strain analysis was performed on the experimental HAADF-STEM image using the geometrical phase analysis (GPA) method. The phase images used for strain calculation were obtained by selecting the (101) and (10-1) reflections from the power spectrum. Supplementary Fig.~1 presents the $\epsilon_{xx}$ strain map at the STO/LSAT interface, where the highest strain is localized at the misfit dislocation cores, appearing as characteristic butterfly-like features.

\subsubsection*{Time-resolved x-ray diffraction}
Time-resolved x-ray diffraction from the Bernina endstation of SwissFEL was taken using 8550 eV x-rays with near-normal incidence on the sample as shown in Fig.~\ref{fig:Fig1}b. Reconstruction of the images to reciprocal space was done using calibration with a 3-d printed pattern to orient the angular coordinates on the 2-d detector. The precise angular offsets and sample orientation were refined using the static diffraction measured with high precision at the Material Science beamline at the Swiss Light Source. Time traces and RSVs were measured at 5~K. The THz pulse with vertical polarization was produced with optical rectification from a LiNbO$_3$ crystal, and the THz field was characterized with electro-optic sampling during the time-resolved x-ray diffraction experiment. 

\subsubsection*{Classical molecular dynamics simulations}
To model the force interactions of the cubic SrTiO$_{3}$ system,  the empirical core-shell model of Li \emph{et al.}~\cite{Li_2021} based on the work of Sepliarsky \emph{et al.}~\cite{Sepliarsky_2005} was used.

To model a cubic SrTiO$_{3}$ thin film geometry a $20\times20\times20$ array of bulk unit-cell SrTiO$_{3}$ (with lattice constant 3.905 Angstroms) was constructed. Fixed periodic boundary conditions where imposed along the in-plane ($x$-$y$) directions and an open boundary was employed along the ($z$) plane normal. This defect-free geometry is referred to as the pristine sample. To introduce the interfacial misfit dislocation array, the bottom $20\times20\times4$ region is changed to an x-y compressed $22\times22\times4$ unit-cell geometry, introducing an addition distorted unit-cell along the $x$ and $y$ directions. In all subsequent simulations (including the pristine geometry), the lowest unit-cell layer is frozen to model the presence of a rigid substrate. Upon relaxation, the introduced lattice mismatch is accommodated by the emergence of two parallel [100] and two parallel [010] dislocations in lower $x$-$y$ plane (Supplementary Fig.~4). This geometry constitutes the sample containing an interfacial misfit dislocation array. The resulting simulation cell has an in-plane periodicity of 15.62 nm with a dislocation spacing of 7.81 nm, and the simulated film thickness is 7.81 nm, which smaller than the experimental sample, which has a dislocation spacing of 40 nm and a thin fim thickness of 35 nm. 

Both the pristine sample and the sample containing misfit dislocations are equilibrated at 300~K using NVT ensemble molecular dynamics. NVE ensemble molecular dynamics is then used to simulate the microscopic response of the in-plane electric field, which is applied to the entire dynamical region along the $[110]$ direction. Each finite temperature MD step was calculated at 0.0001~ps intervals, and every 500 MD steps the atomic configuration saved (simulation time resolution of 0.05~ps).  The experimental in-plane electric field profile was used (Supplementary Fig.~2). To account for the difference in time scales of the wave propagation in the simulated cell compared to experiment, we considered scalings of the in-plane and out-of-plane sizes of the systems. The main text FFT spectra are shown scaled in time according to the difference in in-plane geometries. Supplementary Fig.~5 shows scaling according to the difference in out-of-plane system size. Overall both appear to give a reasonable qualitative match. However, we note that the  the differences in simulation system size may impact the precise identification (or even mix) of the lowest frequency modes of the simulation (with equal in-plane and out-of-plane dimensions), which is different than in the experiment, where there are differences in in-plane and out-of-plane dimensions.

Due to the limited size of the simulation cell, the simulated XRD has a reduced resolution in reciprocal space (Supplementary Fig.~6a-b). Moreover, due to the difference in size of the experimental and simulated supercells of the MD network, a direct comparison of the equivalent points in reciprocal space for the superstructure satellites is not straightforward. For this reason we select a slice in reciprocal space that is similar to that of the experiment, which requires to sample \:$+/-$ 1 plane in reciprocal space to capture the dynamic response of the superstructure required to identify the possible modes present at the dislocations. The simulated time traces are from the sum of the modes from the two planes that are closest to and enclose the experimentally measured reciprocal space plane shown in Supplementary Fig.~6c). The time-dependent atomic strain tensor was obtained using OVITO\cite{Stukowski_2009}. The first (static) frame of the simulation was used as the reference cell for the time-dependent deformation calculation. Supplementary Figs. 12-13 show the strain calculated from the cubic Ti atom sublattice, which gives a clear first coordination sphere in the pair distribution function of 4.7~\AA \:used as the cutoff radius for the calculation. We found no substantial difference in the strain result when considering all atoms, but the relatively smaller Sr displacement and non-cubic O sublattice introduces additional noise. 

\subsubsection*{3-dimensional visualization}
The simulation cell in Supplementary Fig.~4 was visualized in OVITO\cite{Stukowski_2009} and the 3-dimensional RSVs were visualized using ParaView\cite{paraview}.

\section*{Acknowledgements}
We thank Valerio Scagnoli for help with vector visualization discussions and for support for x-ray reflectometry measurements and Bill Pedrini for assistance to measure 3-dimensional RSV at the Materials Science endstation of the Surface Diffraction beamline of the Swiss Light Source. Time-resolved hard x-ray diffraction experiments were performed at the Bernina endstation of ARAMIS branch of SwissFEL at Paul Scherrer Institute. E.S. was supported by the NCCR Materials’ Revolution: Computational Design and Discovery of Novel Materials (NCCR MARVEL No. 182892) from the Swiss National Foundation and the European Union’s Horizon 2020 research and innovation programme under the Marie Sklodowska-Curie Grant Agreement No. 884104 (PSI-FELLOW-III-3i). H.U. was supported by the National Centers of Competence in Research in Molecular Ultrafast Science and Technology (NCCR MUST-No. 51NF40-183615) from the Swiss National Science Foundation and from the European Union’s Horizon 2020 research H. U. also acknowledges the innovation program under the Marie Sklodowska-Curie Grant Agreement No. 801459–FP-RESOMUS. M.Ba. acknowledges support from the Swiss National Science Foundation (Ambizione project, PZ00P2 216089).

\section*{Author contributions}
U.S., E.S., and P.D. conceived of the study. E.S., M.Bu., H.U., R.M., M.San., H.T.L., R.T.W., Y.D., M.Sav., B.L., M.Ba., V.O., S.L.J., and U.S. carried out the experiment. P.M.D. carried out the classical molecular dynamics simulations. M.R. prepared the sample. M.D.R. performed STEM measurements. E.S. analyzed and interpreted the data from experiment and simulations. E.S., wrote the manuscript with U.S. and P.M.D. and with input from all coauthors.

\section*{Competing interests}
The authors declare no competing interests.

\section*{Data and materials availability}
Data will be made available after publication at the Paul Scherrer Institute Data Repository. 

\end{document}


\title[Article Title]{THz-induced phonon mode mixing and collective dynamics in a polar nanolattice}

\author*[1]{\fnm{Elizabeth} \sur{Skoropata}}\email{elizabeth.skoropata@psi.ch}
\author[2]{\fnm{Peter M.} \sur{Derlet}}\email{peter.derlet@psi.ch}
\author[1]{\fnm{Hiroki} \sur{Ueda}}
\author[1]{\fnm{Roman} \sur{Mankowsky}}
\author[1]{\fnm{Mathias} \sur{Sander}}
\author[1]{\fnm{Henrik T.} \sur{Lemke}}
\author[3]{\fnm{Rafael T.} \sur{Winkler}}
\author[1]{\fnm{Yunpei} \sur{Deng}}
\author[1]{\fnm{Milan} \sur{Radovic}}
\author[3]{\fnm{Matteo} \sur{Savoini}}
\author[1]{\fnm{Biaolong} \sur{Liu}}
\author[3]{\fnm{Martina} \sur{Basini}}
\author[3]{\fnm{Vladimir} \sur{Ovuka}}
\author[3]{\fnm{Steven L.} \sur{Johnson}}
\author[4]{\fnm{Marta D.} \sur{Rossell}}
\author[1]{\fnm{Urs} \sur{Staub}}\email{urs.staub@psi.ch}

\affil*[1]{\orgdiv{Center for Photon Science}, \orgname{Paul Scherrer Institute}, \orgaddress{\city{Villigen}, \postcode{5232}, \country{Switzerland}}}

\affil[2]{\orgdiv{Center for Scientific Computing, Theory and Data}, \orgname{Paul Scherrer Institute}, \orgaddress{\city{Villigen}, \postcode{5232}, \country{Switzerland}}}

\affil[3]{\orgdiv{Institute for Quantum Electronics, Physics Department}, \orgname{ETH Zurich}, \orgaddress{\city{Zurich}, \postcode{8092}, \country{Switzerland}}}

\affil[4]{\orgdiv{Electron Microscopy Center}, \orgname{Empa – Swiss Federal Laboratories for Materials Science and Technology}, \orgaddress{\city{D\"ubendorf}, \postcode{8600}, \country{Switzerland}}}

\maketitle

\doublespacing

\vspace{-10mm}
\section*{Supplementary Figures}

\begin{figure}
	\centering
	\includegraphics[width=0.8\textwidth]{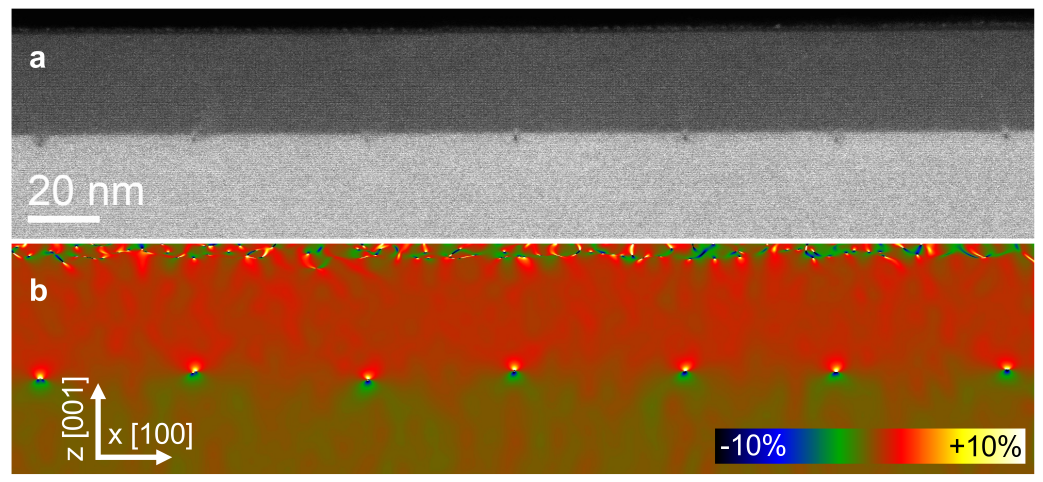}
	\caption{\textbf{a} Cross-sectional scanning transmission electron microscopy (STEM) image and the corresponding in-plane $\epsilon_{xx}$ strain map of the STO/LSAT sample, revealing a periodically ordered array of misfit dislocations at the interface. \textbf{b} The dislocation cores are clearly identifiable in the strain map as characteristic butterfly-like features. The color scale represents strain values between $-10$~\% and $+10$~\%.}
	\label{fig:STEM_strain} 
\end{figure}

\begin{figure}
	\centering
	\includegraphics[width=0.5\textwidth]{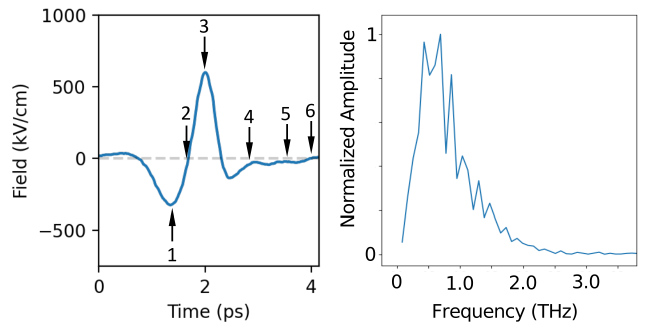} 
	\caption{THz pulse measured with electro-optic sampling during the experiment. Numbers indicate key time points 1 = $t_{\rm{min}}$ (minimum THz field amplitude), 2 = $t_{\rm{rev}}$ (THz field reversal point), 3 =  $t_{\rm{max}}$ (maximum THz field amplitude), 4 = $t_{\rm{off}}$ (immediately after the THz pulse), and 5-6 are selected points after the THz pulse.}
	\label{fig:THz_pulse} 
\end{figure}

\begin{figure}
	\centering
	\includegraphics[width=1.0\textwidth]{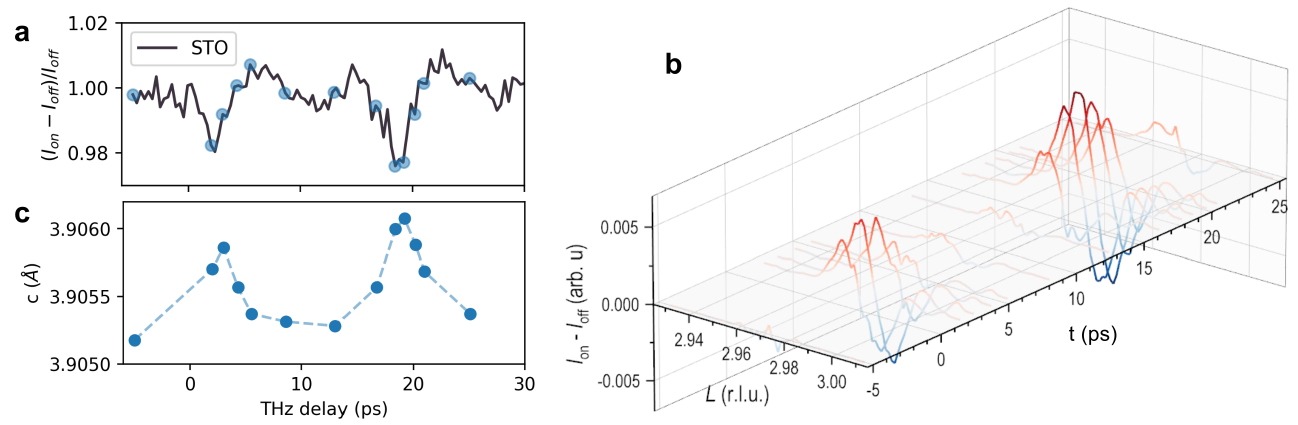}
	\caption{\textbf{a} Time trace of the bulk STO reflection (from Main text Figure~\ref*{fig:Fig1}c) with the points where 3D RSVs were measured indicated in blue. \textbf{b} Intensity contrast for a line cut along the central STO Bragg reflection at (1 -1 $L$) obtained from RSVs taken at various pump-probe delays and \textbf{c} time dependence of the out-of-plane $c$-axis lattice parameter extracted from the $L$ line cuts.}
	\label{fig:xrd_lattice_c} 
\end{figure}

\begin{figure} 
	\centering
	\includegraphics[width=0.9\textwidth]{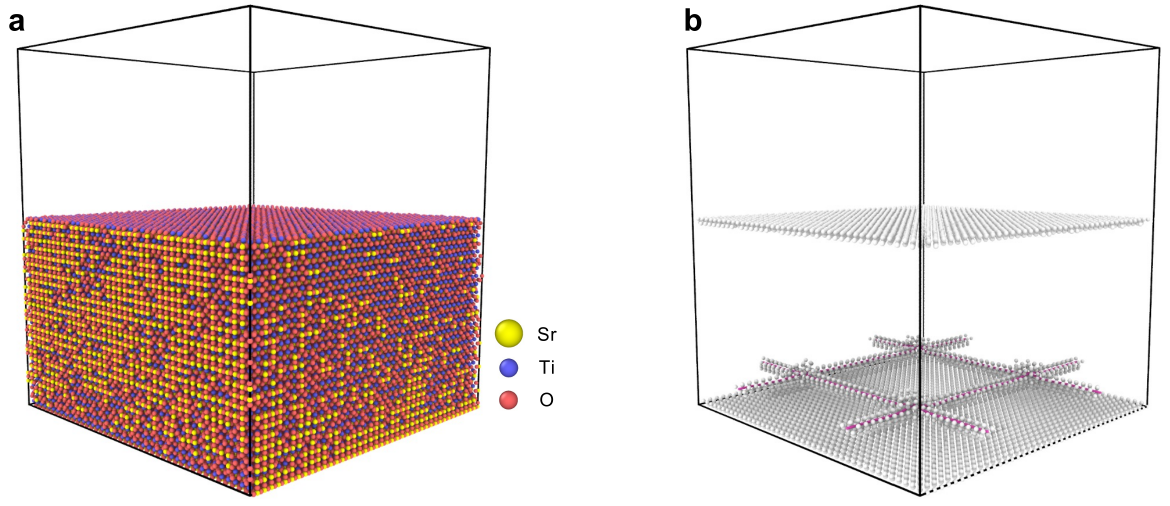} 
	\caption{\textbf{a} Simulation cell used for SrTiO$_3$ with an interfacial misfit dislocation network showing all atoms. \textbf{a} The same simulation cell showing mainly the atoms at the dislocation lines and the film surface.}
	\label{fig:sim_cell} 
\end{figure}

\begin{figure}
	\centering
	\includegraphics[width=0.3\textwidth]{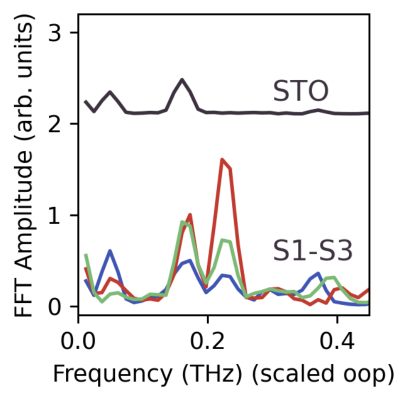} 
	\caption{FFT spectra of the simulated tr-XRD (Main text Figure~\ref*{fig:Fig1}f) using a scaling according to the difference in out-of-plane sizes of the simulation vs. experimental lattice.}
	\label{fig:oop_scaling} 
\end{figure}

\begin{figure}
	\centering
	\includegraphics[width=0.85\textwidth]{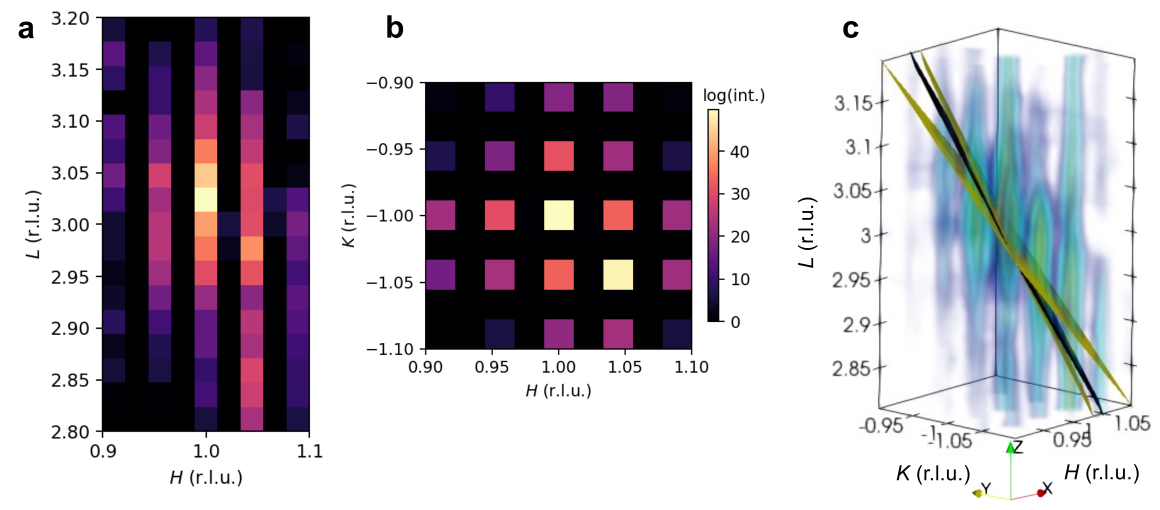} 
	\caption{Projections of the simulated XRD volume \textbf{a} in the $H$-$L$ direction and \textbf{b} in the $H$-$K$ direction, and \textbf{c} 3-dimensional representation. The black plane matches the plane observed in the experimental time traces, and the yellow planes are the two nearest bounding planes that the simulations provide that were used to obtain the simulated XRD time traces.}
	\label{fig:RSV_slices} 
\end{figure}

\begin{figure}
	\centering
	\includegraphics[width=1.0\textwidth]{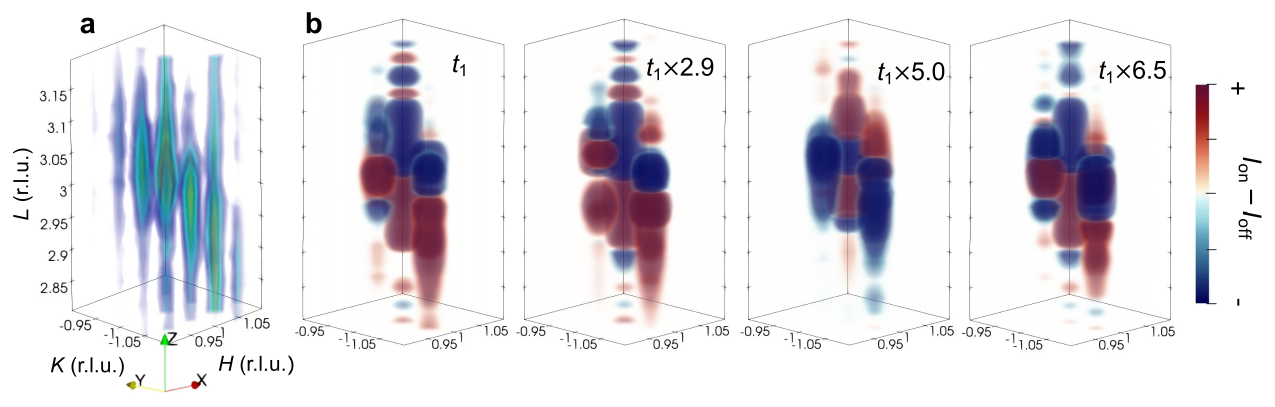} 
	\caption{\textbf{a} Static x-ray RSV from simulation and \textbf{b} simulated XRD contrast ($I_{\rm{on}} - I_{\rm{off}}$) at time points at similar intervals as the experimental RSVs (using the same out-of-plane time-scaling from the difference in simulation cell and experimental thin film used for the FFTs from the time-dependent XRD in Main text Figure~\ref*{fig:Fig1}).}
	\label{fig:RSV_diff} 
\end{figure}

\begin{figure}
	\centering
	\includegraphics[width=0.8\textwidth]{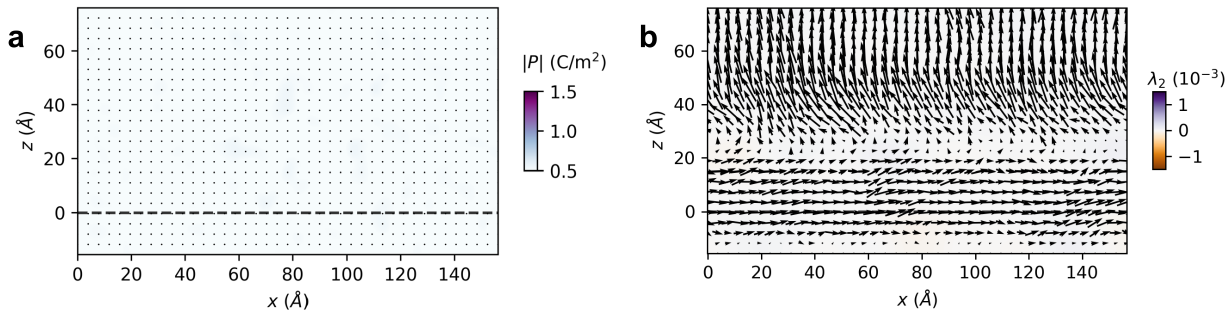} 
	\caption{\textbf{a} Static electric polarization $|P|$ and \textbf{b} $\lambda_2$ of SrTiO$_3$, with intensity scales and time points identical to those of SrTiO$_3$ with dislocations in the main text (Figure~\ref*{fig:Fig4}). The black arrows indicate the O$^{2-}$ displacement. }
	\label{fig:sim_noMD} 
\end{figure}

\begin{figure}
	\centering
	\includegraphics[width=0.6\textwidth]{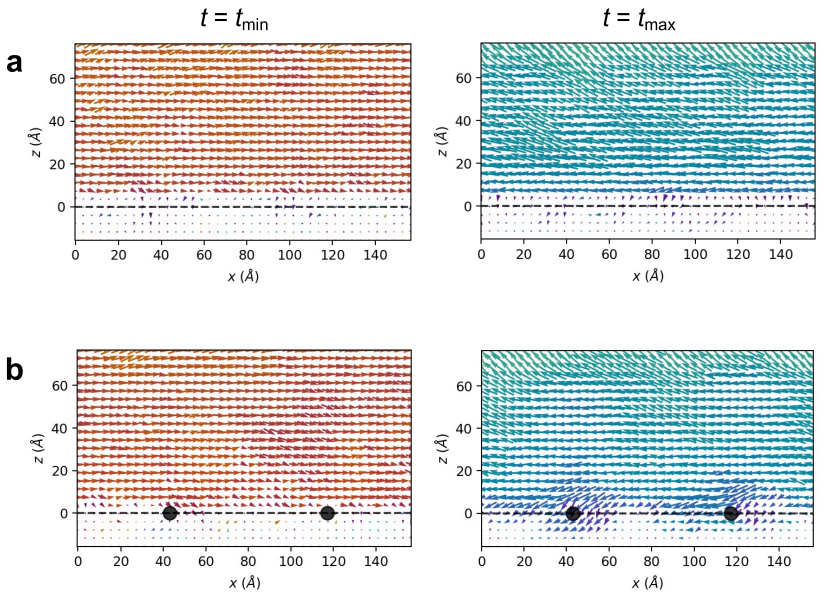} 
	\caption{Mean O$^{2-}$ displacement projected in the (x-z)/(lattice $a$-$c$) crystal planes at the largest negative field of the THz pulse ($t = t_{\rm{min}}$), and at the peak of the THz $E$-field  ($t = t_{\rm{peak}}$) for the simulation cells \textbf{a} of the pristine STO film  \textbf{b} STO with dislocations.}
	\label{fig:THz_O_displacement} 
\end{figure}

\begin{figure}
	\centering
	\includegraphics[width=1.0\textwidth]{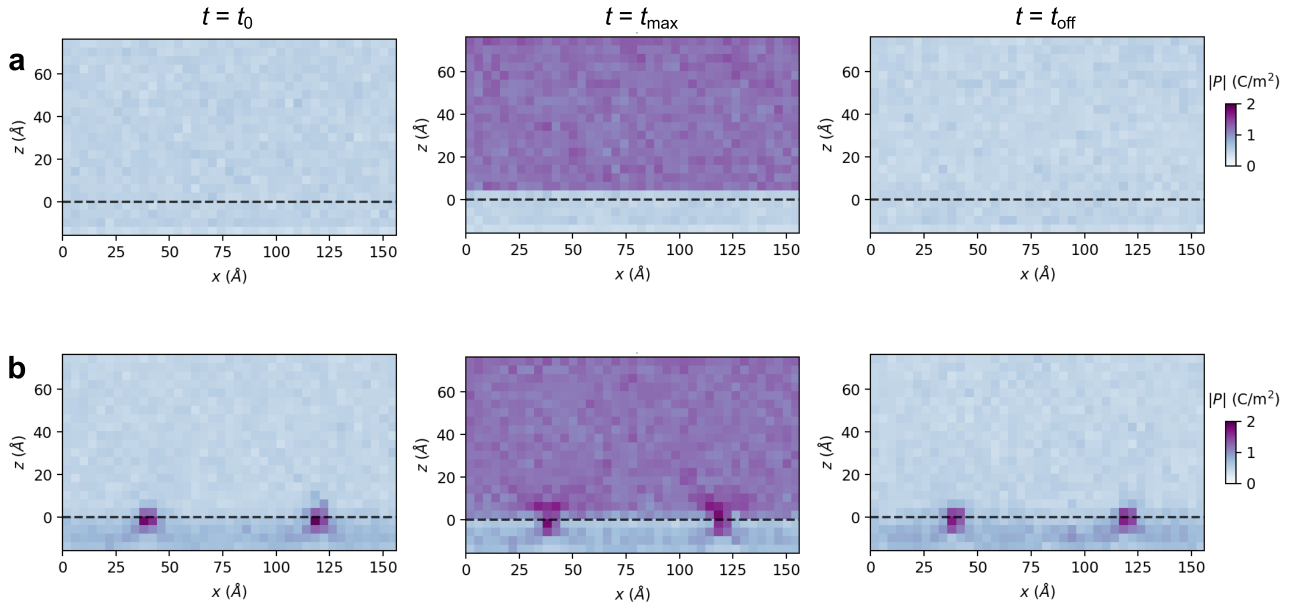} 
	\caption{Projection of the mean magnitude of the electric polarization $|P|$ of \textbf{a} the pristine STO film and \textbf{b} the STO film with interfacial dislocations for the time points $t = t_0$ (before the THz field), $t = t_{\rm{max}}$, and $t = t_{\rm{off}}$ immediately after the THz pulse (from Supplementary Figure~\ref*{fig:THz_pulse}).}
	\label{fig:sim_poln} 
\end{figure}

\begin{figure}
	\centering
	\includegraphics[width=1.0\textwidth]{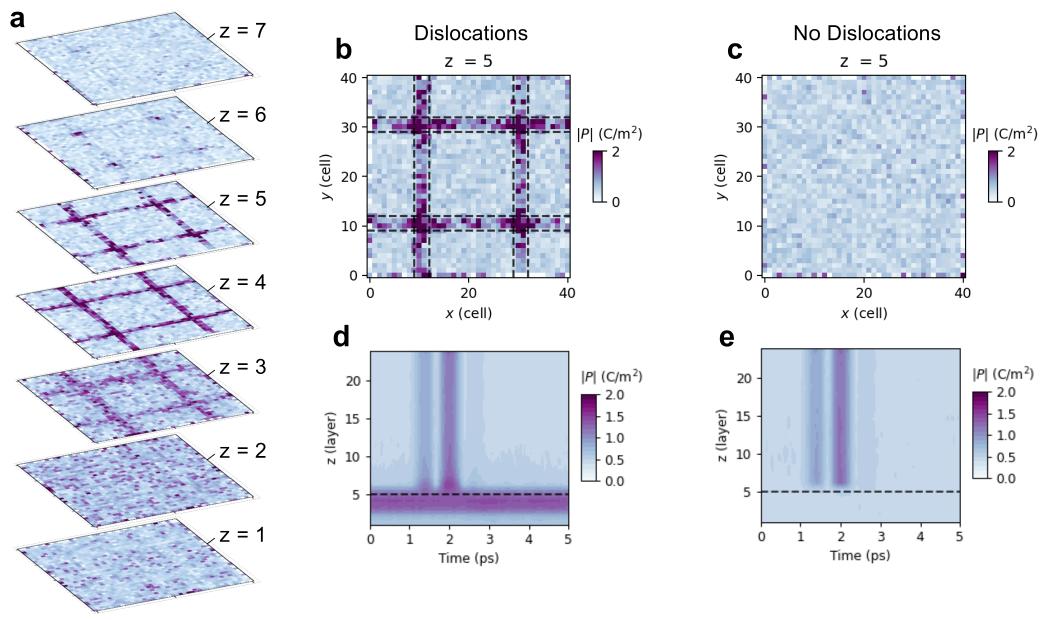} 
	\caption{\textbf{a} Unit-cell layer-by-layer variation of the magnitude of the electric polarization $|P|$ of the STO film with interfacial dislocations. A comparison of $|P|$ immediately above the interface for \textbf{b} STO with interfacial dislocations and \textbf{c} the pristine STO film. Time-dependence of $|P|$ from the interfacial region for \textbf{d} STO with interfacial dislocations and \textbf{e} the pristine STO film. Note that the dashed lines in (b) show the region used to monitor the time variation of $|P|$ for STO with dislocations in (d). Note that the THz pulse is applied from $\sim$1.0 - 2.5~ps. }
	\label{fig:poln_depth} 
\end{figure}

\begin{figure}
	\centering
	\includegraphics[width=1.0\textwidth]{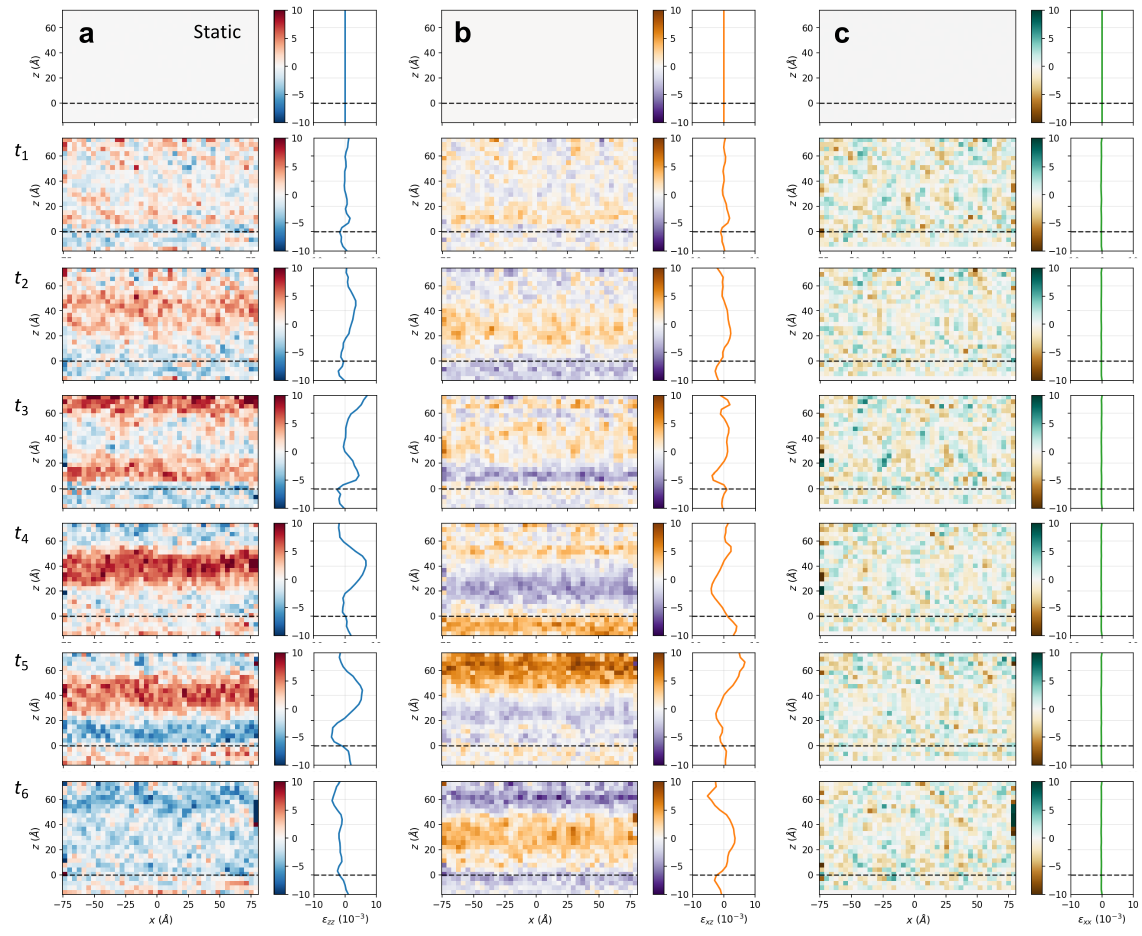} 
	\caption{Strain profile projected in the (x-z)/(lattice b-c) and average depth-dependent strain profiles obtained from simulations of the pristine STO film during the time points in Supplementary Fig.~\ref*{fig:THz_pulse} during and after the THz excitation. Representative components of the strain tensor reflecting \textbf{a} longitudinal $\epsilon_{zz}$, \textbf{b} shear $\epsilon_{xz}$, and \textbf{c} transverse ($\epsilon_{xx}$) components of dynamic strain are shown. Note that we define the strain terminology in the frame of reference provided by the pristine STO film assuming a predominant group velocity in the out-of-plane direction.}
	\label{fig:strain_no_dislocations} 
\end{figure}

\begin{figure}
	\centering
	\includegraphics[width=1.0\textwidth]{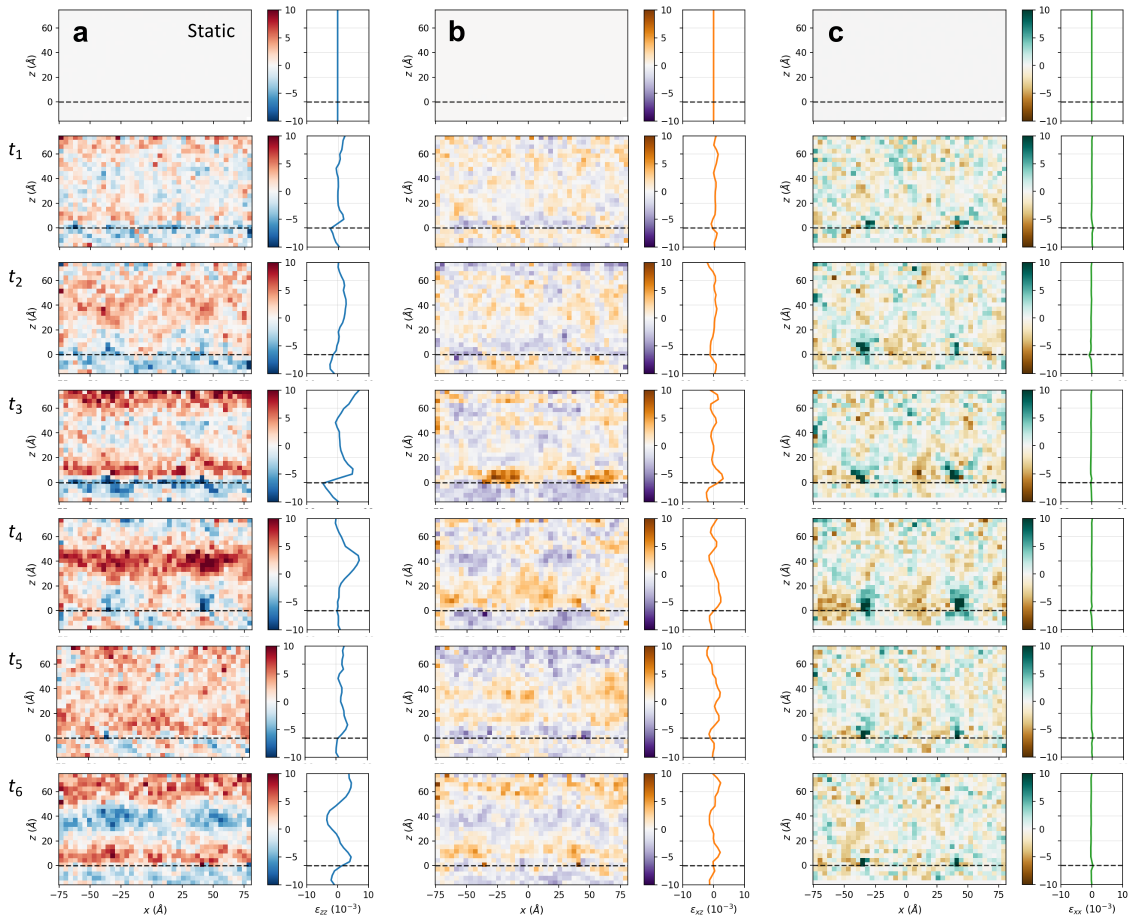} 
	\caption{Strain profile projected in the (x-z)/(lattice b-c) and average depth-dependent strain profiles obtained from simulations of the STO film with dislocations during the time points in Supplementary Fig.~\ref*{fig:THz_pulse} during and after the THz excitation. Representative components of the strain tensor reflecting \textbf{a} longitudinal $\epsilon_{zz}$, \textbf{b} shear $\epsilon_{xz}$, and \textbf{c} transverse ($\epsilon_{xx}$) components of dynamic strain are shown. Note that we define the strain terminology for STO with dislocations within the same frame of reference established by the the pristine STO film to facilitate comparison between the two simulation cells.}
	\label{fig:strain_dislocations} 
\end{figure}

\clearpage

\section*{Movie Captions}
\paragraph{Caption for Movie S1.}
\textbf{Atomic displacement induced by THz pulse in an STO thin film} 
Animated version of Main text Figures~\ref*{fig:Fig3}-\ref*{fig:Fig4} of the x-z (lattice $a$-$c$) projected mean displacement of O$^{2-}$ ions in a pristine STO thin film during and after a THz pulse obtained from molecular dynamic simulations.

\paragraph{Caption for Movie S2.}
\textbf{Atomic displacement induced by THz pulse in STO with interfacial dislocations}
Animated version of Main text Figures~\ref*{fig:Fig3}-\ref*{fig:Fig4} of the x-z (lattice $a$-$c$) projected mean displacement of O$^{2-}$ ions in an STO film with interfacial dislocations during and after a THz pulse obtained from molecular dynamic simulations.